\documentclass[%
 aip,
sor,
 amsmath,amssymb,
preprint,%
]{revtex4-1}

\usepackage{graphicx}
\usepackage{dcolumn}
\usepackage{bm}
\usepackage{amsmath}
\usepackage{subfigure}
\usepackage{xcolor}

\begin{document}

\preprint{APS/123-QED}

\title{Stress fluctuations and shear thickening in dense granular suspensions}

\author{Qin Xu}
\affiliation{James Franck Institute, University of Chicago, IL, 60637 USA}
\affiliation{Department of Materials, ETH Zurich, 8093 Zurich, Switzerland}
\author{Abhinendra Singh}
\affiliation{James Franck Institute, University of Chicago, IL, 60637 USA}
\affiliation{Pritzker School of Molecular Engineering, University of Chicago, IL, 60637 USA}
\author{Heinrich M. Jaeger}
\affiliation{James Franck Institute, University of Chicago, IL, 60637 USA}
\affiliation{Department of Physics, University of Chicago, IL, 60637 USA}

\begin{abstract}
We experimentally investigate the rheology and stress fluctuations of granules densely suspended in silicone oil.
We find that both thickening strength and stress fluctuations significantly weaken with oil viscosity $\eta_0$. Comparison of our rheological results to the Wyart-Cates model for describing different dynamic jamming states suggests a transition from frictional contacts to lubrication interactions as $\eta_0$ increases. To clarify the contribution from viscous interactions to the rheology, we systematically measure stress fluctuations in various flow states. Reduction of stress fluctuations with $\eta_0$ indicates that a strong lubrication layer greatly inhibits force correlations among particles. Measuring stress fluctuations in the strong shear thickening regime, we observe a crossover from asymmetric Gamma to symmetric Gaussian distributions and associated with it a decrease of lateral (radial) correlation length $\xi$ with increasing shear rate.

\end{abstract}

\maketitle

\section{Introduction}
Concentrated particle suspensions are found in both nature and industry~\cite{coussot2017mudflow,benbow1993paste}.
Dense granular suspensions are mixtures of non-Brownian particles (diameter $\geq 1\mu m$) and viscous liquids at high volume fractions, and can exhibit complex flow behaviors far from Newtonian fluids~\cite{mewis_colloidal_2011,Brown2014,Denn_2018}.
Because of the large solid fraction $\phi$, direct contacts between grains are crucial in governing the rheology \cite{Brown2014, Mari2015, Seto2013}.
 Similar to  dry granular materials, dense granular suspensions can dilate against a boundary under flow \cite{Brown2012}. 
As the dilation is counteracted by confinement, particles are likely to rearrange and create dynamic force networks via frictional contacts.
As a critical volume fraction is approached, a steep jump in flow resistance with shear rate can be observed, a phenonmenon called discontinuous shear thickening (DST) \cite{Wyart2014, Fall2008, Brown2009, Brown2014,Seto2013}.
However, viscous interactions can also be significant in suspensions as  grains are close to each other at high volume fractions \cite{Stickel2005,Maiti2014}. For instance, the contribution of lubrication forces to shear thickening has been discussed extensively. Depending on the system, viscous coupling has been found to enhance \cite{Bender1996, Cheng2011}, weaken \cite{Xu2014} or not affect shear thickening \cite{Lin2015}. 
%

Recent studies~\cite{Seto2013,Fernandez_2013,Mari_2014,Lin2015,Guy_2015,Mari2015,Ness_2016, Clavaud_2017,Comtet_2017} have shown that strong shear thickening is a stress-driven transition from a lubricated ``frictionless" to an unlubricated ``frictional" state. 
A critical stress $\tau_c$ is required to squeeze out the lubrication layer present between the particles and bring them into frictional contacts.
This critical stress could originate from the repulsive force due to steric (e.g. due to adsorbed polymer) or electrostatic stabilization as well as Brownian force, since all these forces essentially keep the particles apart as long as the shear stresses are smaller than a certain threshold level.
%
%
The frictional forces are influenced by particle roughness~\cite{Lootens2005, Fernandez_2013, Hsu_2018} or chemical surface interactions~\cite{james2018interparticle}.
The behavior can be thus divided into two regimes that are essentially rate independent, i.e., states in which the stresses are linear with strain rate.
One is the low-stress regime ($\tau \ll \tau_c$), where particles interact through lubrication forces, with the rheology diverging at the frictionless jamming point $\phi_{\rm J}^0$.
%
In this case, the viscosity is affected by the long-ranged attractive or short--ranged repulsive 
forces~\cite{Brown_2010, Singh_2019}.
The second regime is at high stresses ($\tau \gg \tau_c$), where almost all particles come into direct, frictional contacts, and the viscosity (and also the normal stresses)
diverge at a volume fraction $\phi_{\rm J}^\mu < \phi_{\rm J}^0$ that depends on the interparticle friction coefficient $\mu$.
With increasing stress $\tau$, the crossover between these two rate independent regimes results in the shear thickening behavior.
Thus, a complete description of the dynamics of dense granular suspensions requires the understanding of both frictional and viscous interactions.

Prior works have indicated that stresses in a dense particulate suspension can exhibit strong fluctuations\cite{Lootens2003,Lootens2005,Dasan2002}. 
Here we exploit the fact that the statistical features of globally measured stress 
fluctuations may
help to understand the contributions of different local interactions to the underlying rheology.
%
%
%
%
For dry granular materials, stress fluctuations and their distributions are of importance in identifying the frictional contacts and micro-structures in different mechanical states, such as jamming, yielding, strain stiffening and shear banding \cite{MillerBrian;OHernCorey;Behringer1996,Howell1999a,Blair2001,Corwin2005,Corwin2008,singh2014effect}.
Sheared dense suspensions share the importance of frictional interactions with dry granular media. However, the link between stress 
fluctuations
and time-averaged 
flow properties, such as the effective suspension viscosity that emerges from the interplay of lubrication and friction has not been investigated systematically in the
DST regime. 

We show quantitatively that, as the solvent viscosity increases,  the lubrication interactions greatly reduce frictional contacts, and we observe significant weakening of both shear thickening and stress fluctuations. Comparing stress distributions for increasing sample sizes, we systematically measure spatial correlations the strong shear thickening regime. Our results show that stress correlations are also weakened by viscous interactions. The work presented here not only demonstrates a different approach to control thickening but also reveals how local statistic properties, in this case stress fluctuations, are linked to the global rheology of dense granular suspensions.

\section{Methods: experimental setup and model}
\subsection{Experimental setup}
Two different hard particle suspensions are studied in our experiments: glass spheres and ZrO$_2$ beads suspended in silicone oil. Glass beads (average diameter 22 $\pm$ 5 $\mu$m) disperse
well in silicone oil and are chosen as a typical shear
thickener with low yield stress. For 
 imaging the boundary profile at the free surface of the suspension we use ZrO$_2$ particles that are larger (200 $\pm$ 10 $\mu$m in diameter) and less reflective than glass. To adjust the viscous interactions in the suspensions, the viscosity ($\eta_0$) of the silicone oil is varied from $20$ cSt to $1.8\times10^4$ cSt by tuning the molecular weight of the polymer chains from $10^3$ to $10^6$ g/mole. For this range of solvent viscosities, particle inertia can be neglected.

Flow properties are meausred with an Anton Paar PCR301 rheometer using a parallel-plate geometry. Care is taken to make sure no fluid extended outside the parallel plates so the grains are
confined to the space between the plates by liquid surface tension. For rheological measurements of steady state flow, we shear suspensions by gradually ramping up shear rate $\dot\gamma$. For each $\dot\gamma$, we measure shear stress $\tau$ and viscosity is calculated by the ratio ${\tau}/{\dot \gamma}$. For measurements of stress fluctuations, we keep shear rate $\dot{\gamma}$ unchanged and measure shear stress evolving with time, $\tau(t)$.

\subsection{Model}
Below, we summarize the current phenomenological approach to model the rheology of dense suspensions.
The basic assumption ~\cite{Wyart2014,Ness_2016,Singh_2018} is that
 the relative viscosity $\eta_{\rm r}=\eta(\phi,\tau)/\eta_0$ is in distinct stress-independent states both at low and high stresses.
The viscosity $\eta_{\rm r}$ is thus simply a function of the packing fraction $\phi$ at low ($\tau \ll \tau_c$) and high ($\tau \gg \tau_c$) stress states and can be expressed as
   \begin{subequations}\label{eq:eta_phi}
    \begin{equation}\label{eq:etaL_phi}
    \eta_{\rm r}^{\rm L} (\phi) =  (1-\phi/\phi_{\rm J}^0)^{-2}
   \end{equation}
   \begin{equation}
   \label{eq:etaH_phi}
    \eta_{\rm r}^{\rm H}(\phi,\mu)=  (1-\phi/\phi_{\rm J}^\mu)^{-2}~,
   \end{equation}
   \end{subequations}
   where $\phi_{\rm J}^0$ and $\phi_{\rm J}^\mu$ denote the jamming volume fractions 
   for $\mu = 0$ (lubricated, frictionless state) and nonzero values of $\mu$ (frictional state), respectively.

Next, to introduce the stress dependence on viscosity and to capture the transition between the lubrication dominated and frictionally dominated states, a stress-dependent jamming volume fraction
is specified using an expression similar to that proposed by Wyart and Cates \citep{Wyart2014} 
   \begin{equation}\label{phi_str}
  \phi_{\rm m}({\tau}) = \phi_{\rm J}^\mu  f({\tau}) + \phi_{\rm J}^0 [ 1-f({\tau}) ]~,
   \end{equation}
   where $f({\tau})$ denotes the fraction of particle interactions for which the shear force exceeds the critical force $F_c$ to bring the particles into direct frictional contact.
 Using~\eqref{eq:eta_phi}, and \eqref{phi_str} the stress dependent viscosity $\eta_{\rm r} (\phi, {\tau})$ can be expressed as:
       \begin{equation}\label{eq:eta_str_phi}
     \eta_{\rm r} (\phi,{\tau}) = [1 - \phi/\phi_{\rm m} ({\tau})]^{-2} .
   \end{equation}

Before we discuss our results it is important to mention that the experiments performed here are for particles sufficiently dense that they cannot be matched in specific density with the suspensing liquid  and will settle under gravity. The onset stress for strong shear thickening is therefore not controlled by overcoming electrostatic repulsive forces, as in many colloidal systems. Instead, the minimum stress to bring particles into direct contact is associated with entraining them  into the shear flow.  This leads to the following picture.
%
The particles are first dispersed within the rheometer and sediment in the absence of any applied shear.
As shear is applied, the suspension overcomes the gravitational force, it fluidizes and exhibits thinning behavior as was probed in detail in our previous study~\cite{Xu2014}.
In the fluidized state, the suspension now behaves like a density matched non--Brownian (or Brownian) suspension at low stress wherein the lubricated interaction between particles essentially lead to divergence of viscosity at $\phi_{\rm J}^0$.
As the shear rate is ramped up, the contact stresses are able to break the  lubrication layer between particles and interparticle friction then leads to shear thickening 
as observed in other dense suspensions~\cite{Brown2014,Denn_2018}.
%
\begin{figure}[t]
\begin{center}
\includegraphics[width=80mm]{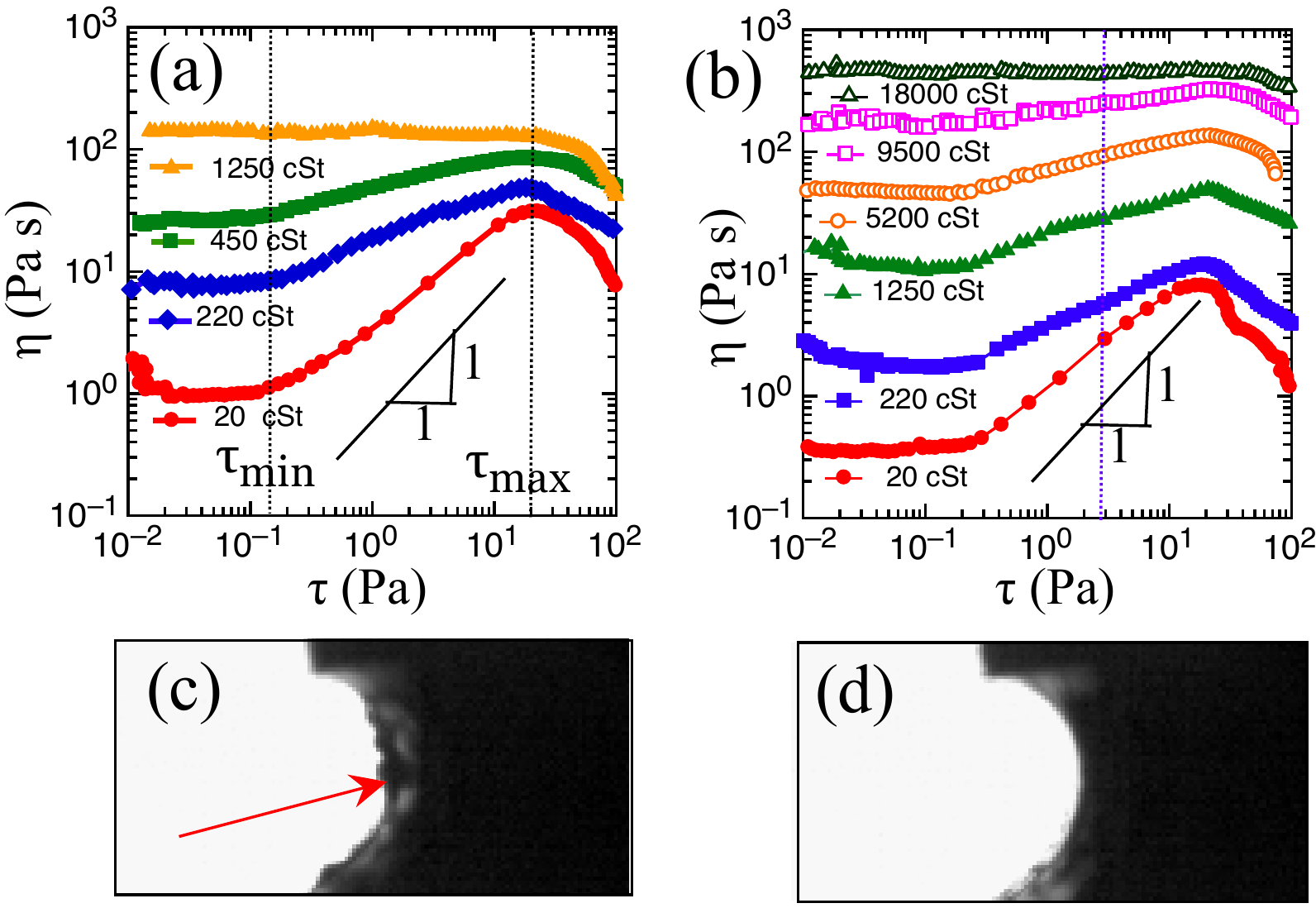}
\end{center}
	\caption{{\bf Rheology.} Flow curves of suspension viscosity ($\eta$) vs. stress ($\tau$) for (a) 22$\mu$m glass beads in oil with volume fraction $\phi=55\%$ and (b) 203$\mu$m ZrO$_2$ particles in oil with volume fraction $\phi=56\%$ for different oil viscosities. (c) and (d) are images taken from suspensions containing 20 cSt and $9.5\times 10^3$ cSt silicone oil under a given shear stress $\tau=2.8$ Pa, respectively. Red arrow: roughness of the interface caused by  frustrated dilation.}
\end{figure}
\section{Results}
\subsection{Rheological analysis}
\begin{figure*}[t]
\begin{center}
\includegraphics[width=120mm]{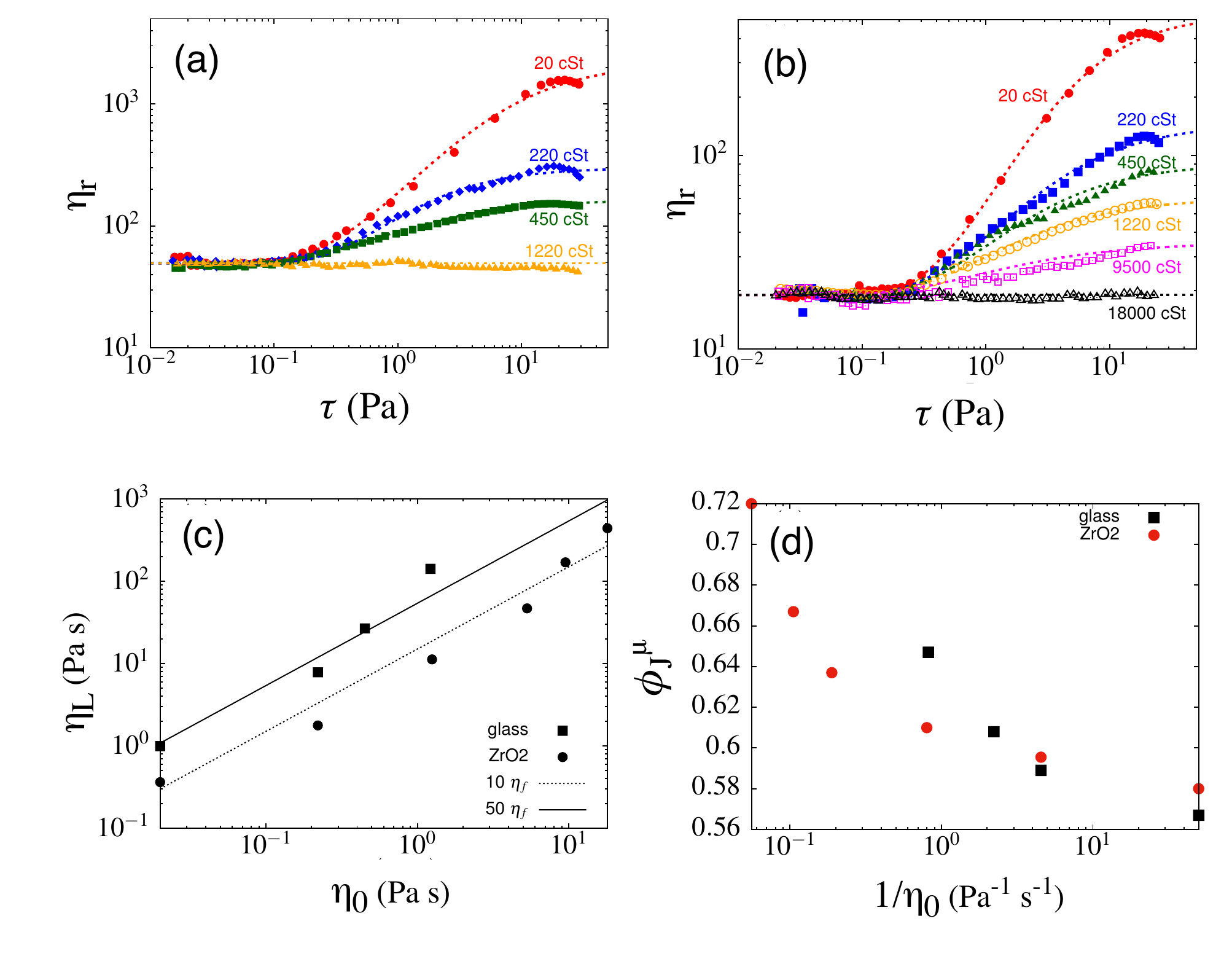}
\end{center}
\caption{
{\bf Rheological analysis.}
Relative viscosity $\eta_{\mathrm r}$ as a function of stress $\tau$ for (a) glass-oil (for $\phi =55\%$) and (b) ZrO$_2$-oil (for $\phi =56\%$) suspensions. Symbols are experimental data with different solvent viscosity $\eta_0$ as indicated with lines that represent a fit to Eq.~\eqref{eq:eta_str_phi}.
(c) Viscosity of the low stress (lubricated) state plotted as a function of the solvent viscosity $\eta_0$.
(d) Frictional jamming volume fraction $\phi_{\mathrm J}^\mu$ plotted as a function of inverse solvent viscosity $1/\eta_0$ for glass--oil and ZrO$_2$-oil suspensions.
}
\label{fig:rheo_analysis}
\end{figure*}
Figure 1(a) shows flow curves for dense glass-oil suspensions in varying viscosities of silicone oil for a constant volume fraction $\phi= 55\%$.
For $\eta_0=20, 220, 450$ cSt, shear thickening is observed within the same stress range between $\tau_{min}$ and $\tau_{max}$. The onset stress $\tau_{min} \approx 0.14$ Pa indicates the minimum stress required to rearrange the particle configurations by overriding the gravity \cite{Brown2012}, and the ending stress $\tau_{max}\approx 20$ Pa is set by the maximum Laplace pressure acting on the particles at the suspension-air interface without moving the contact line \cite{Xu2014}. Given the fact that both $\tau_{min}$ and $\tau_{max}$ are independent of solvent viscosity, the two bounds of shear thickening are not varied by the viscous dissipation in the suspensions.  The degree of the shear thickening, however, changes dramatically with $\eta_0$.

Taking the suspension with $\eta_0=20$ cSt (red solid circles in Fig.1(a)) as an example, we observe an approximate $\eta\sim\tau$ relation, which indicates an
abrupt, almost continuous increase of viscosity at a critical shear rate $\dot{\gamma}_c$. The value of $\dot{\gamma}_c$ for the viscosity jump is given by the slope of $\eta({\tau})$.  As we increase $\eta_0$, we observe a crossover from strong shear thickening to stress independent flow behavior. As shown in Fig. 1a, shear thickening weakens for 220 cSt (blue diamonds) and 450 cSt (green squares), and vanishes at $\eta_0 = 1250$ cSt (orange triangles).
%
A similar crossover is also observed for ZrO$_2$-oil suspensions (Fig. 1(b)). In this case, a much larger oil viscosity ($\eta_0\approx 1.8\times 10^4$ cSt) is required to obscure shear thickening and achieve a rate independent behavior. In such a highly viscous limit, each stress data point is taken by averaging 10$^4$ results over a period of 2 hours to ensure steady state.

We image the interface of ZrO$_2$ suspensions {\em in-situ} as rheological measurements are performed. The suspensions are lit from behind as a camera focuses on the interface. The profile is highlighted by the contrast between white background and dark suspensions. As shown in Figs. 1(c)\&(d), we compare the surface profiles between $\eta_0=20$ cSt (Fig. 1c) and $\eta_0=9.5\times10^3$ cSt (Fig. 1d) under a same shear stress $\tau=2.8$ Pa in shear thickening regime. For $\eta_0=20$ cSt, interface is roughened by the protruded particles as shear thickening occurs, which is consistent with frustrated dilation \cite{Brown2012}. For $\eta_0=9.5\times10^3$ cSt, however, the suspension interface remains smooth and liquid-like. 

At low stresses ($\tau < \tau_{min}$), since the lubrication film is present between the particles, the
viscosity of suspension $\eta_{\mathrm L}$ is expected to scale with the solvent viscosity $\eta_0$.
The low stress viscosity $\eta_{\mathrm L}$ for both glass--oil and ZrO$_2$-oil suspensions is displayed
as a function of $\eta_0$ in Fig.~\ref{fig:rheo_analysis}c, and we recover a linear relation between the them as $\eta \propto \eta_0$.
Some of the deviations from the linear trend could arise from the existence of a yield stress or could be associated with the difficulty of mixing of solid particles homogeneously in highly viscous fluid at high volume fractions $\phi \sim 0.55$.
Figures~\ref{fig:rheo_analysis}a and b display the relative viscosity $\eta_{\rm r}$ as a function of stress $\tau$ for glass--oil and ZrO$_2$-oil suspensions for various values of solvent viscosity $\eta_0$.
In these plots we have used the relations displayed in Fig.~\ref{fig:rheo_analysis}c to extract the relative viscosity $\eta_{\rm r}$, to collapse the viscosity data at low stresses.
We observe that the high--stress frictional viscosity $\eta_{\mathrm H}$ decreases with increase in the solvent viscosity, and also the extent of shear thickening decreases with increase in solvent viscosity $\eta_0$.
The decrease in $\eta_{\mathrm H}$ suggests that higher molecular weight makes it difficult to bring the particles into direct, frictional contact.
%
%

Simulations have demonstrated that, for a given volume fraction, the viscosity of the frictional state and hence the extent of thickening decreases with the decrease in interparticle friction coefficient $\mu$~\cite{Mari_2014, Singh_2018}, while the onset stress for shear thickening is independent of $\mu$.
The flow--curves reported in Figs.~\ref{fig:rheo_analysis}a and b for different $\eta_0$ clearly show a very similar behavior.
To test the hypothesis that direct frictional contacts and hence contact force networks drive the observed changes with different molecular weight solvent, the friction--based model of Refs.~\cite{Wyart2014, Singh_2018} is fit to the shear--thickening portion
($\tau_{\mathrm{min}} < \tau < \tau_{\mathrm{max}}$) of all viscosity curves.
As the last missing ingredient in Eq.~\eqref{eq:eta_str_phi}, the fraction of particles in direct, frictional contacts $f(\tau)$ is expressed as a $f(\tau) = \exp(-\tau^*/\tau)$, where $\tau^*$ is the critical stress for particles to overcome the critical force.

The friction-based model (Eq.~\eqref{eq:eta_str_phi}) very well describes the shear thickening in samples with variable $\eta_0$ (Fig.~\ref{fig:rheo_analysis}a,b) using fixed $\phi_J^0$ and $\tau^*$ and variable $\phi_J^\mu$ for glass bead and ZrO$_2$ suspensions.
The resulting parameters $\phi_J^0$ = 0.65 and $\tau^* = 0.45$ for glass-oil, and $\phi_J^0$ = 0.72  and $\tau^* = 0.55$ for ZrO$_2$--oil suspensions were found to be independent of $\eta_0$. Previous work for hard--sphere colloidal suspensions reported a value of $\phi_J^0 = 0.71$ ~\cite{Royer_2016}.
%

Since $\tau^*$ is found to be independent of the solvent viscosities $\eta_0$, this indicates that polymers with different molecular weight do not affect the critical force needed to bring particles into frictional contact. 
Whereas $\phi_J^0$ and $\tau^*$ are found to be insensitive to $\eta_0$, the striking result here is that the frictional jamming point $\phi_J^\mu$ shows a pronounced increase with $\eta_0$.
This increase in $\phi_J^\mu$ is consistent with the thought that an increase in polymer molecular weight inhibits the formation of frictional contacts and the development of a system--spanning frictional contact network.

\begin{figure*}[t]
\begin{center}
\includegraphics[width=140mm]{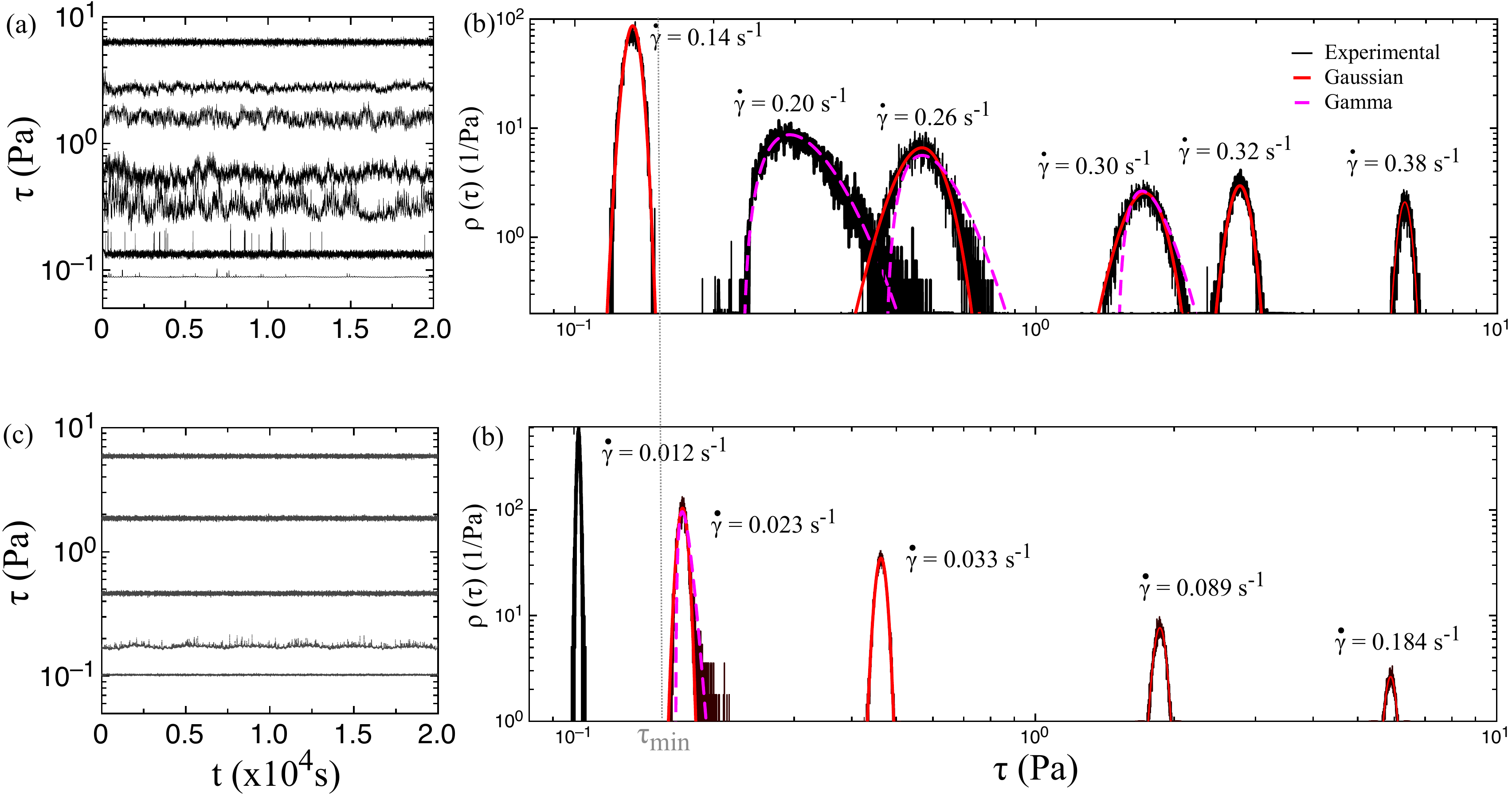}
\end{center}
	\caption{{\bf Shear stress fluctuations.} (a) Time evolution of stress for shear rates at $\dot{\gamma}=$ 0.09, 0.14,0.20, 0.26, 0.30,0.32,0.38 s$^{-1}$ (from the bottom to top), measured for glass beads in 20 cSt silicone oil. (b) Probability density distribution  functions $\rho(\tau)$ as from the fluctuation data in (a) for $\dot{\gamma}=$ 0.14,0.20, 0.26, 0.30,0.32,0.38 s$^{-1}$. Solid black lines are the experimental results. Red solid lines and pink dashed lines are  fits to Gaussian and Gamma distributions. The grey dashed line points at the onset of shear thickening ($\tau_{min}$). Similarly, panel (c) shows the time evolution of stress for shear rates at $\dot{\gamma}$= 0.012, 0.023, 0.033, 0.089, 0.184s$^{-1}$ (from the bottom to top)  for glass beands in 220 cSt silicone oil, and the corresponding distributions are shown in panel (d).}
\end{figure*}
\subsection{Stress fluctuations}
To further validate our hypothesis that weakening of  frictional contacts with increasing $\eta_0$ drives the observed change in rheology, we now turn our attention to the stress fluctuations.
Previous studies~\cite{boersma_1990, Lootens2003,rathee_2017} have shown that dense hard--sphere suspensions can exhibit pronounced fluctuations and that the magnitude of these fluctuations depends on the closeness to the DST transition.
However none of these studies have related the fluctuations to interparticle forces.
%

We apply constant shear to glass-oil suspensions to measure the temporal fluctuations of the shear stress $\tau(t)$. In this section, we focus on the comparison of results with $\eta_0$ = 20 and 220 cSt. For the higher solvent viscosity, stress fluctuations are hardly visible. Figure 3(a), for instance, displays the plots of $\tau(t)$ for $\eta_0$ = 20 cSt over a period of $2\times 10^4$ seconds. From bottom to top, each trace corresponds to a different applied shear rate as indicated. When $\dot\gamma\geq 0.20 $s$^{-1}$, shear stresses are in the range of shear thickening,  $\tau_{min}<{\tau} (t,\dot\gamma)<\tau_{max}$. For the top six traces of $\tau (t,\dot\gamma)$ in panel (a), we plot the probability density function, $\rho(\tau)$, in panel (b).

The profiles of $\rho(\tau)$ evolve with applied shear rate and shear stress. For $\dot\gamma=0.14$ s$^{-1}$, the corresponding shear stress $\tau\vert_{\dot\gamma=0.14 s^{-1}}$ is below shear thickening onset $\tau_{min}$. As are result, we find that $\rho(\tau)$ is well approximated by a Gaussian distribution with a standard deviation set by the stress resolution of our instruments ($\sim 0.01$ Pa). As soon as the shear stress passes $\tau_{min}$, the 
fluctuations of shear stress grow significantly and the distribution starts to deviate
from Gaussian. At $\dot\gamma=0.20$ s$^{-1}$, for example, the shear stress is only slightly above $\tau_{min}$ but $\rho(\tau)$ is no longer symmetric. Instead, $\rho(\tau)$ can be fitted well by a second-order Gamma function
\begin{equation}
\rho(\tau)=\frac{(\tau-\tau_0)^2}{2\tau_1^3}  \exp(-\frac{(\tau-\tau_0)}{\tau_1}),
\end{equation}
where $\tau_0$ $(\approx 0.23 \pm 0.03)$ Pa and $\tau_1$ $(\approx 0.031\pm0.008)$ Pa are fitting parameters that characterize the mean and
the effective width, respectively. Equation (4) is consistent with the results of the so-called q-model, which predicts stress fluctuations from a force chain network connected through direct contacts \cite{Coppersmith1996}. Such distribution is also found for the normal stress of sheared dry granular media \cite{MillerBrian;OHernCorey;Behringer1996} where frictional interactions dominate particle contacts. However, the distribution becomes less asymmetric as we increase the shear rate, $\dot\gamma=0.26, 0.30, 0.32, 0.38$ s$^{-1}$, driving the shape of $\rho(\tau)$ back to Gaussian again, but with a larger standard deviation ($\sim 10^0$ Pa). Note that the rheometer shear stress
noise is significantly smaller ($\sim 10^{-2}$ Pa).

Fluctuations of shear stress greatly reduce for higher oil viscosity. Figures 2c and d show the plots of $\tau(t)$ and $\rho(t)$ for $\eta_0=220$cSt. Clearly, the magnitude of fluctuations decreases significantly. Except possibly for $\dot\gamma=0.023$ s$^{-1}$, $\rho(\tau)$ can be well fitted by Gaussian distributions for most of shear rates and measuring radii. So, why do we see different stress distributions  for different flow states in shear thickening regime?   

\begin{figure*}[t]
\begin{center}
\includegraphics[width=100mm]{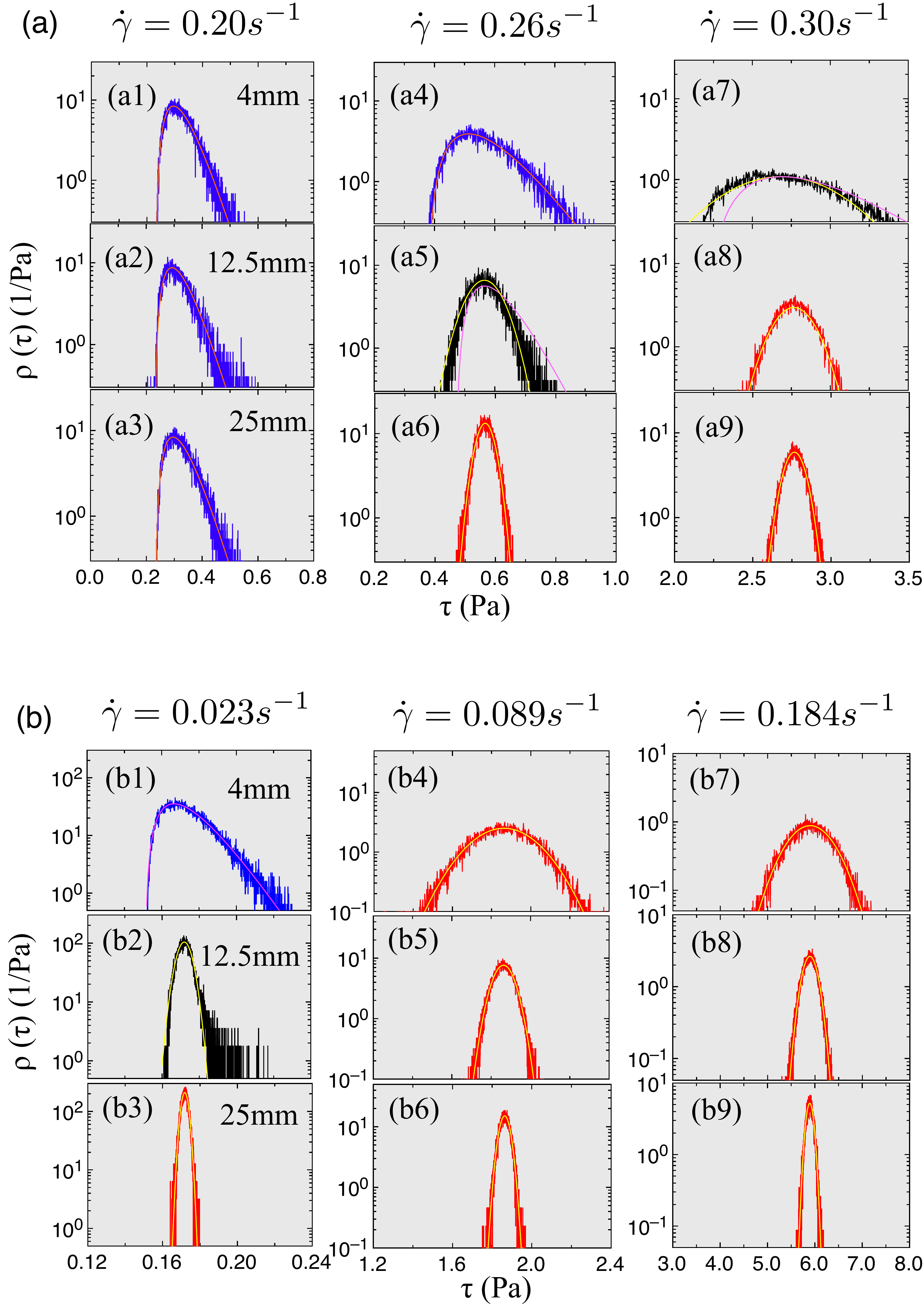}
\end{center}
	\caption{{\bf System size dependence.} Plots of $\rho(\tau)$ for (a) $\eta_0=20$ cSt at different shear rates (columns, left to right: 0.20s$^{-1}$, 0.26s$^{-1}$ and 0.30s$^{-1}$) and (b) $\eta_0=220$ cSt at different shear rates (columns, left to right: 
	0.023s$^{-1}$,
	0.089s$^{-1}$,  and 0.184s$^{-1}$). From  top to  bottom: plate size $R_s=4, 12.5$ and $25$ mm. }
\end{figure*}

One likely scenario is connected to the  stress-dependent contact forces. For $\tau < \tau_{min}$ all contacts are lubricated and the distribution is Gaussian simply due to experimental measurement uncertainty. In the beginning shear thickening ($\tau\sim\tau_{min}$), frictional force chains start to emerge and span to gap between. As shear rate increases in shear thickening regime, more force chains are lined up between plates. It is important to note that Eq.(4) is based on an ensemble average over individually normalized distributions in a static packing of particles. In the steady-state shear experiments, however, the force networks are  constantly evolving. Therefore, if the measurement window becomes longer than the reconfiguration time, and if subsequent realizations of the contact force network are uncorrelated, the experiments will record averages over distributions  as in Eq.(4). This will lead to a Gaussian-like distribution due to the Central Limit Theorem (CLT), in line with what we observe in Fig.3. When more independent force chains are built up as we are ramping up $\dot\gamma$, $\rho(\tau)$ becomes more close to a Gaussian distribution.

We can further test this scenario by exploring spatial correlations. We do that by measuring the shear stress fluctuations in samples that have a different lateral (radial) extent, which we prepare by varying the volume of suspension placed between plates of a given, fixed spacing.  If the sample subdivides into several uncorrelated regions we would expect a reduction in the standard deviation of fluctuations around their mean. Figure 4 shows the stress fluctuation probability density $\rho(\tau)$ for different sample radii $R$ for both low (a) and high (b) solvent viscosity. Each column shows a different average shear stress, corresponding to  $\dot\gamma$ =0.20 , 0.26
and 0.30 s (left to right). In these panels, we use various colors to indicate different distributions. Blue and red curves are used to present $\rho(\tau)$ that can be fitted well by Gamma and Gaussian distributions, respectively. For $\rho(\tau)$ partly fitted by either Gammma or Gaussian functions, we use black.

In Fig. 4(a), we observe a crossover from Gamma to Gaussian distributions with increasing $R$. Meanwhile, the standard deviation of $\rho(\tau)$ decreases. The results are indeed similar to the averaging effects in the case of ramping up $\dot{\gamma}$. At low shear rate $\dot\gamma=0.20$ s$^{-1}$ however, $\rho(\tau)$ is independent of $R$. The distribution plots in Fig. 4(a1) - (a3) can be nicely fitted with a single set of parameters $\tau_0 = 0.23 \pm 0.05$ Pa and $\tau_1 = 0.031\pm0.009$ Pa as $R$ increases from 4 to 25 mm. Therefore, at low stresses in the early stage of shear thickening, all force chains are highly correlated across the whole sample.

As solvent viscosity increases to $\eta_0=220$ cSt in Fig. 4b, most of the distribution functions in the panel are nicely fitted by Gaussian distributions (the fitted results are plotted in yellow lines). The only asymmetric Gamma distribution is found at low shear rate $\dot\gamma=0.023$ s$^{-1}$ with small sample radius $R=4$ mm. As long as $R$ or $\dot\gamma$ increases, $\rho(\tau)$ is quickly averaged into a Gaussian distribution. In contrast to the low-viscosity case, now the standard deviation greatly depends on sample radius. As $R$ increases from 4 to 25 mm, the contact area  between the suspension and the plates increases by a factor around 40. According to the CLT, averaging results of independent variables will reduce the standard deviation by a factor of $\sqrt{40}\approx 6.3$, which is consistent with the plots of $\rho(\tau)$ (for example comparing Fig. 4(b4) to (b6)). The results implies that viscous interactions in suspensions weaken stress correlations among particles.

We can make this quantitative by defining a spatial correlation length $\xi$ along the lateral (i.e.e, radial) direction by looking into the change of standard deviation with $R$. This standard deviation of the fluctuation distributions is defined as $\sigma = \sqrt{\overline{(\tau(t) - \tau_{average})^2}}$. For a Gaussian
distribution the mean stress $\tau_{average}$ and  $\sigma$ follow the usual definitions, while for the Gamma distribution in Eq.(4)  $\tau_{average} = \tau_0 + 3 \tau_1$ and  $\sigma = \sqrt{3} \tau_1$.
When  $R\gg \xi$, the suspension includes $N$ sub-regions, each having a certain characteristic contact area with the top and bottom plates. As the sample radius R increases, the number N of sub-regions scales as the total contact area, which is proportional to $R^2$, and consequently  $\sigma\sim1/\sqrt{N}\sim 1/R$. On the other hand, when $R \ll \xi$, the stress across entire sample is spatially correlated such that $\rho(\tau)$ and $\sigma$ are independent of $R_s$. Figure 5(a) provides an example to extract the value of $\xi$. For $\eta_0=20$ cSt,  $\sigma$ is plotted against  $R_s$ for $\dot\gamma=0.32$ s$^{-1}$. It is clear that  $\sigma$ remains  constant at small radius and decreases with $R_s$ at large radius. From the plot, we can clearly define a transition point as the correlation length scale $\xi\approx4.3$ mm.

\begin{figure}[t]
\begin{center}
\includegraphics[width=120mm]{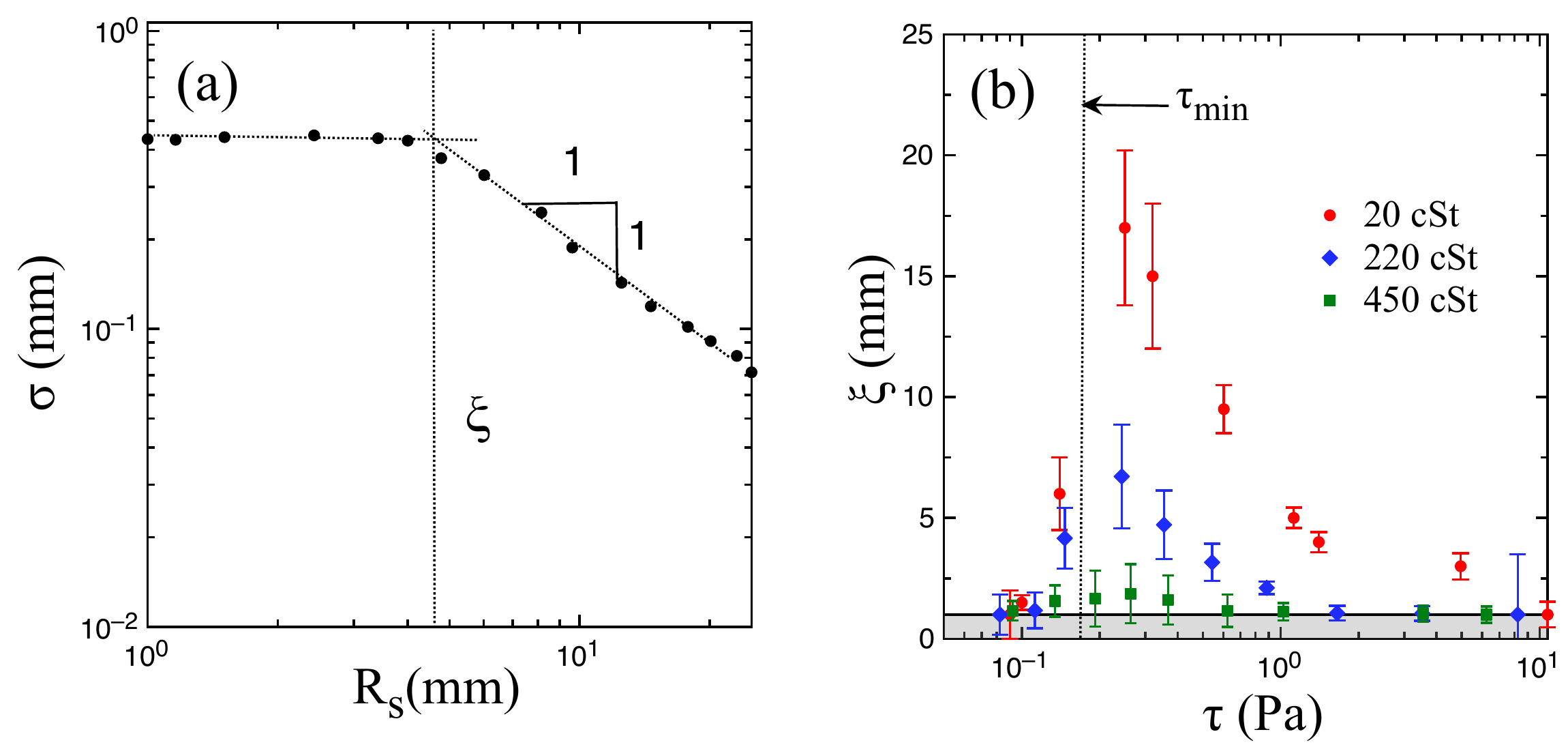}
\end{center}
	\caption{{\bf Spatial correlations}. (a)  Standard deviation $\sigma$ is plotted against sample radius $R_s$ at $\dot{\gamma}=0.32$ s$^{-1}$. The transition point from constant to $\sigma\sim 1/R_s$ decay indicates the value of $\xi$. (b) The plot of $\xi$ vs. $\bar\tau$ at $\eta_0 = 20, 220$ and $450$ cSt. The dashed line is the onset stress of the shear thickening ($\tau_{min}$). The grey region labels the lower limit of our sample size.}
\end{figure}

Figure 5(b) plots the lateral correlation length scale , $\xi$, against the mean value of $\tau$ at a given shear rate for different solvent viscosities. For $\eta_0=20$ cSt (solid red points), $\xi$ gradually increases with $\tau$ when $\tau<\tau_{min}$. Beyond the shear thickening onset $\tau_{min}$, $\xi$ increases dramatically. Here the precision of our standard deviation measurements is not sufficient to tell
whether $\xi$ is possibly divergent near $\tau_{min}$. Nevertheless, the sudden increase in $\xi$ in approaching shear thickening regime signals the formation of a network of frictional contacts spanning the measuring gap, consistent with our expectation regarding the mechanisms driving DST. However, as the shear rate further increases into the shear thickening regime, $\xi$ decreases with applied shear stress, as already implied quantitatively in the plots of stress distributions $\rho(\tau)$. As shown in Figure 3(b), the transition from Gamma to Gaussian distributions is a signature of an averaging effect as we ramp up the shear rate. The result suggests a increasing number of independent stress-carrying sub-regions is formed as shear thickening strengthens.  A similar trend of $\xi$ is also observed for higher solvent viscosity. For $\eta_0=220$ and $450$ cSt (solid blue and green points), although the values of $\xi$ are reduced significantly compared to $\eta_0=20$ cSt, the plot shows a quantitatively similar dependence on $\tau$. The values of $\xi$ increase with $\tau$ as $\tau<\tau_{min}$, then decrease with $\tau$ beyond $\tau_{min}$, and hence peak near the shear thickening onset.  

Similar results have been observed in previous works as well. For example, Lootens \cite{Lootens2003} observed a crossover from asymmetric to Gaussian distributions for stress fluctuations of dense colloidal suspensions near the jamming point. In fact, if we interpret strong shear thickening and in particular DST as intermittent jamming behavior, where frictional force chain networks continually form, break up and reconfigure, we would require strong correlations only in the vertical direction that connects the shearing plates. Only for the ultimate limit of a solid-like, shear jammed state might a more isotropic, system-spanning network by expected ~\cite{Bi2011}. Our results thus suggest that shear-induced force networks in DST regime may not be isotropic. We expect further experiments to clarify the issue. 

\section{Conclusions}
We show experimentally that viscous interactions in dense granular suspensions can greatly weaken the shear thickening strength and decrease stress fluctuations. First, by comparing steady state rheology measurements for different viscosities $\eta_0$ to the Wyart-Cates model, we demonstrate that the frictional jamming point $\phi_J^\mu$ increases with solvent viscosity. The result is equivalent to  decreasing thickening strength by lowering the interparticle friction. Second, by systematically measuring stress fluctuations in the shear thickening regime for different $\eta_0$, we quantify how viscous interactions reduce force correlations among particles and weaken the formation of a network of frictional contact forces.

Both results indicate that the solvent viscosity decreases frictional contacts in suspensions. This is likely due to a significant increase of the molecular weight of the polymers as we increase $\eta_0$. For instance, as the viscosity of silicone oil increases by 50 times, the molecular weight can increase orders of magnitude correspondingly. Such long polymer chains can easily be entangled and form a robust lubrication layer as particles are sheared against each other. Our results demonstrate the competition between frictional and viscous interactions in dense suspensions,and they introduce a new approach to  control this competition and  thereby tune strength of shear thickening.

\section{Acknowledgments}
We thank Ivo R. Peters and Carlos S. Orellana
for many useful  discussions. This work was supported by the the US Army Research Office through Grant No. W911NF-16-1-0078 and the Chicago MRSEC through NSF Grant No. DMR-1420709. A.S. acknowledges support from the Center for Hierarchical Materials Design (CHiMaD).
\section{References}
\bibliography{reference}

\begin{thebibliography}{42}%
\makeatletter
\providecommand \@ifxundefined [1]{%
 \@ifx{#1\undefined}
}%
\providecommand \@ifnum [1]{%
 \ifnum #1\expandafter \@firstoftwo
 \else \expandafter \@secondoftwo
 \fi
}%
\providecommand \@ifx [1]{%
 \ifx #1\expandafter \@firstoftwo
 \else \expandafter \@secondoftwo
 \fi
}%
\providecommand \natexlab [1]{#1}%
\providecommand \enquote  [1]{``#1''}%
\providecommand \bibnamefont  [1]{#1}%
\providecommand \bibfnamefont [1]{#1}%
\providecommand \citenamefont [1]{#1}%
\providecommand \href@noop [0]{\@secondoftwo}%
\providecommand \href [0]{\begingroup \@sanitize@url \@href}%
\providecommand \@href[1]{\@@startlink{#1}\@@href}%
\providecommand \@@href[1]{\endgroup#1\@@endlink}%
\providecommand \@sanitize@url [0]{\catcode `\\12\catcode `\$12\catcode
  `\&12\catcode `\#12\catcode `\^12\catcode `\_12\catcode `\%12\relax}%
\providecommand \@@startlink[1]{}%
\providecommand \@@endlink[0]{}%
\providecommand \url  [0]{\begingroup\@sanitize@url \@url }%
\providecommand \@url [1]{\endgroup\@href {#1}{\urlprefix }}%
\providecommand \urlprefix  [0]{URL }%
\providecommand \Eprint [0]{\href }%
\providecommand \doibase [0]{http://dx.doi.org/}%
\providecommand \selectlanguage [0]{\@gobble}%
\providecommand \bibinfo  [0]{\@secondoftwo}%
\providecommand \bibfield  [0]{\@secondoftwo}%
\providecommand \translation [1]{[#1]}%
\providecommand \BibitemOpen [0]{}%
\providecommand \bibitemStop [0]{}%
\providecommand \bibitemNoStop [0]{.\EOS\space}%
\providecommand \EOS [0]{\spacefactor3000\relax}%
\providecommand \BibitemShut  [1]{\csname bibitem#1\endcsname}%
\let\auto@bib@innerbib\@empty
\bibitem [{\citenamefont {Coussot}(2017)}]{coussot2017mudflow}%
  \BibitemOpen
  \bibfield  {author} {\bibinfo {author} {\bibfnamefont {P.}~\bibnamefont
  {Coussot}},\ }\href@noop {} {\emph {\bibinfo {title} {Mudflow rheology and
  dynamics}}}\ (\bibinfo  {publisher} {Routledge},\ \bibinfo {year}
  {2017})\BibitemShut {NoStop}%
\bibitem [{\citenamefont {Benbow}\ and\ \citenamefont
  {Bridgwater}(1993)}]{benbow1993paste}%
  \BibitemOpen
  \bibfield  {author} {\bibinfo {author} {\bibfnamefont {J.}~\bibnamefont
  {Benbow}}\ and\ \bibinfo {author} {\bibfnamefont {J.}~\bibnamefont
  {Bridgwater}},\ }\bibfield  {title} {\enquote {\bibinfo {title} {Paste flow
  and extrusion},}\ }\href@noop {} {\  (\bibinfo {year} {1993})}\BibitemShut
  {NoStop}%
\bibitem [{\citenamefont {Mewis}\ and\ \citenamefont
  {Wagner}(2011)}]{mewis_colloidal_2011}%
  \BibitemOpen
  \bibfield  {author} {\bibinfo {author} {\bibfnamefont {J.}~\bibnamefont
  {Mewis}}\ and\ \bibinfo {author} {\bibfnamefont {N.~J.}\ \bibnamefont
  {Wagner}},\ }\href@noop {} {\emph {\bibinfo {title} {Colloidal {Suspension}
  {Rheology}}}}\ (\bibinfo  {publisher} {Cambridge University Press},\ \bibinfo
  {year} {2011})\BibitemShut {NoStop}%
\bibitem [{\citenamefont {Brown}\ and\ \citenamefont
  {Jaeger}(2014)}]{Brown2014}%
  \BibitemOpen
  \bibfield  {author} {\bibinfo {author} {\bibfnamefont {E.}~\bibnamefont
  {Brown}}\ and\ \bibinfo {author} {\bibfnamefont {H.~M.}\ \bibnamefont
  {Jaeger}},\ }\bibfield  {title} {\enquote {\bibinfo {title} {{Shear
  thickening in concentrated suspensions: phenomenology, mechanisms and
  relations to jamming.}}}\ }\href {\doibase 10.1088/0034-4885/77/4/046602}
  {\bibfield  {journal} {\bibinfo  {journal} {Reports on progress in physics}\
  }\textbf {\bibinfo {volume} {77}},\ \bibinfo {pages} {046602} (\bibinfo
  {year} {2014})}\BibitemShut {NoStop}%
\bibitem [{\citenamefont {Denn}, \citenamefont {Morris},\ and\ \citenamefont
  {Bonn}(2018)}]{Denn_2018}%
  \BibitemOpen
  \bibfield  {author} {\bibinfo {author} {\bibfnamefont {M.~M.}\ \bibnamefont
  {Denn}}, \bibinfo {author} {\bibfnamefont {J.~F.}\ \bibnamefont {Morris}}, \
  and\ \bibinfo {author} {\bibfnamefont {D.}~\bibnamefont {Bonn}},\ }\bibfield
  {title} {\enquote {\bibinfo {title} {Shear thickening in concentrated
  suspensions of smooth spheres in newtonian suspending fluids},}\ }\href@noop
  {} {\bibfield  {journal} {\bibinfo  {journal} {Soft Matter}\ }\textbf
  {\bibinfo {volume} {14}},\ \bibinfo {pages} {170--184} (\bibinfo {year}
  {2018})}\BibitemShut {NoStop}%
\bibitem [{\citenamefont {Mari}\ \emph {et~al.}(2015)\citenamefont {Mari},
  \citenamefont {Seto}, \citenamefont {Morris},\ and\ \citenamefont
  {Denn}}]{Mari2015}%
  \BibitemOpen
  \bibfield  {author} {\bibinfo {author} {\bibfnamefont {R.}~\bibnamefont
  {Mari}}, \bibinfo {author} {\bibfnamefont {R.}~\bibnamefont {Seto}}, \bibinfo
  {author} {\bibfnamefont {J.~F.}\ \bibnamefont {Morris}}, \ and\ \bibinfo
  {author} {\bibfnamefont {M.~M.}\ \bibnamefont {Denn}},\ }\bibfield  {title}
  {\enquote {\bibinfo {title} {{Discontinuous shear thickening in Brownian
  suspensions by dynamic simulation}},}\ }\href {\doibase
  10.1073/pnas.1515477112} {\bibfield  {journal} {\bibinfo  {journal}
  {Proceedings of the National Academy of Sciences}\ }\textbf {\bibinfo
  {volume} {112}},\ \bibinfo {pages} {15326--15330} (\bibinfo {year}
  {2015})}\BibitemShut {NoStop}%
\bibitem [{\citenamefont {Seto}\ \emph {et~al.}(2013)\citenamefont {Seto},
  \citenamefont {Mari}, \citenamefont {Morris},\ and\ \citenamefont
  {Denn}}]{Seto2013}%
  \BibitemOpen
  \bibfield  {author} {\bibinfo {author} {\bibfnamefont {R.}~\bibnamefont
  {Seto}}, \bibinfo {author} {\bibfnamefont {R.}~\bibnamefont {Mari}}, \bibinfo
  {author} {\bibfnamefont {J.~F.}\ \bibnamefont {Morris}}, \ and\ \bibinfo
  {author} {\bibfnamefont {M.~M.}\ \bibnamefont {Denn}},\ }\bibfield  {title}
  {\enquote {\bibinfo {title} {{Discontinuous Shear Thickening of Frictional
  Hard-Sphere Suspensions}},}\ }\href {\doibase 10.1103/PhysRevLett.111.218301}
  {\bibfield  {journal} {\bibinfo  {journal} {Physical Review Letters}\
  }\textbf {\bibinfo {volume} {111}},\ \bibinfo {pages} {218301} (\bibinfo
  {year} {2013})}\BibitemShut {NoStop}%
\bibitem [{\citenamefont {Brown}\ and\ \citenamefont
  {Jaeger}(2012)}]{Brown2012}%
  \BibitemOpen
  \bibfield  {author} {\bibinfo {author} {\bibfnamefont {E.}~\bibnamefont
  {Brown}}\ and\ \bibinfo {author} {\bibfnamefont {H.~M.}\ \bibnamefont
  {Jaeger}},\ }\bibfield  {title} {\enquote {\bibinfo {title} {{The role of
  dilation and confining stresses in shear thickening of dense suspensions}},}\
  }\href {\doibase 10.1122/1.4709423} {\bibfield  {journal} {\bibinfo
  {journal} {Journal of Rheology}\ }\textbf {\bibinfo {volume} {56}},\ \bibinfo
  {pages} {875} (\bibinfo {year} {2012})}\BibitemShut {NoStop}%
\bibitem [{\citenamefont {Wyart}\ and\ \citenamefont
  {Cates}(2014)}]{Wyart2014}%
  \BibitemOpen
  \bibfield  {author} {\bibinfo {author} {\bibfnamefont {M.}~\bibnamefont
  {Wyart}}\ and\ \bibinfo {author} {\bibfnamefont {M.~E.}\ \bibnamefont
  {Cates}},\ }\bibfield  {title} {\enquote {\bibinfo {title} {{Discontinuous
  Shear Thickening without Inertia in Dense Non-Brownian Suspensions}},}\
  }\href {\doibase 10.1103/PhysRevLett.112.098302} {\bibfield  {journal}
  {\bibinfo  {journal} {Physical Review Letters}\ }\textbf {\bibinfo {volume}
  {112}},\ \bibinfo {pages} {098302} (\bibinfo {year} {2014})}\BibitemShut
  {NoStop}%
\bibitem [{\citenamefont {Fall}\ \emph {et~al.}(2008)\citenamefont {Fall},
  \citenamefont {Huang}, \citenamefont {Bertrand}, \citenamefont {Ovarlez},\
  and\ \citenamefont {Bonn}}]{Fall2008}%
  \BibitemOpen
  \bibfield  {author} {\bibinfo {author} {\bibfnamefont {A.}~\bibnamefont
  {Fall}}, \bibinfo {author} {\bibfnamefont {N.}~\bibnamefont {Huang}},
  \bibinfo {author} {\bibfnamefont {F.}~\bibnamefont {Bertrand}}, \bibinfo
  {author} {\bibfnamefont {G.}~\bibnamefont {Ovarlez}}, \ and\ \bibinfo
  {author} {\bibfnamefont {D.}~\bibnamefont {Bonn}},\ }\bibfield  {title}
  {\enquote {\bibinfo {title} {{Shear Thickening of Cornstarch Suspensions as a
  Reentrant Jamming Transition}},}\ }\href {\doibase
  10.1103/PhysRevLett.100.018301} {\bibfield  {journal} {\bibinfo  {journal}
  {Physical Review Letters}\ }\textbf {\bibinfo {volume} {100}},\ \bibinfo
  {pages} {018301} (\bibinfo {year} {2008})}\BibitemShut {NoStop}%
\bibitem [{\citenamefont {Brown}\ and\ \citenamefont
  {Jaeger}(2009)}]{Brown2009}%
  \BibitemOpen
  \bibfield  {author} {\bibinfo {author} {\bibfnamefont {E.}~\bibnamefont
  {Brown}}\ and\ \bibinfo {author} {\bibfnamefont {H.}~\bibnamefont {Jaeger}},\
  }\bibfield  {title} {\enquote {\bibinfo {title} {{Dynamic Jamming Point for
  Shear Thickening Suspensions}},}\ }\href {\doibase
  10.1103/PhysRevLett.103.086001} {\bibfield  {journal} {\bibinfo  {journal}
  {Physical Review Letters}\ }\textbf {\bibinfo {volume} {103}},\ \bibinfo
  {pages} {086001} (\bibinfo {year} {2009})}\BibitemShut {NoStop}%
\bibitem [{\citenamefont {Stickel}\ and\ \citenamefont
  {Powell}(2005)}]{Stickel2005}%
  \BibitemOpen
  \bibfield  {author} {\bibinfo {author} {\bibfnamefont {J.~J.}\ \bibnamefont
  {Stickel}}\ and\ \bibinfo {author} {\bibfnamefont {R.~L.}\ \bibnamefont
  {Powell}},\ }\bibfield  {title} {\enquote {\bibinfo {title} {{Fluid Mechanics
  and Rheology of Dense Suspensions}},}\ }\href {\doibase
  10.1146/annurev.fluid.36.050802.122132} {\bibfield  {journal} {\bibinfo
  {journal} {Annual Review of Fluid Mechanics}\ }\textbf {\bibinfo {volume}
  {37}},\ \bibinfo {pages} {129--149} (\bibinfo {year} {2005})}\BibitemShut
  {NoStop}%
\bibitem [{\citenamefont {Maiti}\ and\ \citenamefont
  {Heussinger}(2014)}]{Maiti2014}%
  \BibitemOpen
  \bibfield  {author} {\bibinfo {author} {\bibfnamefont {M.}~\bibnamefont
  {Maiti}}\ and\ \bibinfo {author} {\bibfnamefont {C.}~\bibnamefont
  {Heussinger}},\ }\bibfield  {title} {\enquote {\bibinfo {title} {{Rheology
  near jamming: The influence of lubrication forces}},}\ }\href {\doibase
  10.1103/PhysRevE.89.052308} {\bibfield  {journal} {\bibinfo  {journal}
  {Physical Review E}\ }\textbf {\bibinfo {volume} {89}},\ \bibinfo {pages}
  {052308} (\bibinfo {year} {2014})}\BibitemShut {NoStop}%
\bibitem [{\citenamefont {Bender}\ and\ \citenamefont
  {Wagner}(1996)}]{Bender1996}%
  \BibitemOpen
  \bibfield  {author} {\bibinfo {author} {\bibfnamefont {J.}~\bibnamefont
  {Bender}}\ and\ \bibinfo {author} {\bibfnamefont {N.}~\bibnamefont
  {Wagner}},\ }\bibfield  {title} {\enquote {\bibinfo {title} {{Reversible
  shear thickening in monodisperse and bidisperse colloidal dispersions}},}\
  }\href
  {http://scitation.aip.org/content/sor/journal/jor2/40/5/10.1122/1.550767}
  {\bibfield  {journal} {\bibinfo  {journal} {Journal of Rheology}\ }\textbf
  {\bibinfo {volume} {40(5)}},\ \bibinfo {pages} {899--916} (\bibinfo {year}
  {1996})}\BibitemShut {NoStop}%
\bibitem [{\citenamefont {Cheng}\ \emph {et~al.}(2011)\citenamefont {Cheng},
  \citenamefont {McCoy}, \citenamefont {Israelachvili},\ and\ \citenamefont
  {Cohen}}]{Cheng2011}%
  \BibitemOpen
  \bibfield  {author} {\bibinfo {author} {\bibfnamefont {X.}~\bibnamefont
  {Cheng}}, \bibinfo {author} {\bibfnamefont {J.~H.}\ \bibnamefont {McCoy}},
  \bibinfo {author} {\bibfnamefont {J.~N.}\ \bibnamefont {Israelachvili}}, \
  and\ \bibinfo {author} {\bibfnamefont {I.}~\bibnamefont {Cohen}},\ }\bibfield
   {title} {\enquote {\bibinfo {title} {{Imaging the Microscopic Structure of
  Shear Thinning and Thickening Colloidal Suspensions}},}\ }\href {\doibase
  10.1126/science.1207032} {\bibfield  {journal} {\bibinfo  {journal}
  {Science}\ }\textbf {\bibinfo {volume} {333}},\ \bibinfo {pages} {1276--1279}
  (\bibinfo {year} {2011})}\BibitemShut {NoStop}%
\bibitem [{\citenamefont {Xu}\ \emph {et~al.}(2014)\citenamefont {Xu},
  \citenamefont {Majumdar}, \citenamefont {Brown},\ and\ \citenamefont
  {Jaeger}}]{Xu2014}%
  \BibitemOpen
  \bibfield  {author} {\bibinfo {author} {\bibfnamefont {Q.}~\bibnamefont
  {Xu}}, \bibinfo {author} {\bibfnamefont {S.}~\bibnamefont {Majumdar}},
  \bibinfo {author} {\bibfnamefont {E.}~\bibnamefont {Brown}}, \ and\ \bibinfo
  {author} {\bibfnamefont {H.~M.}\ \bibnamefont {Jaeger}},\ }\bibfield  {title}
  {\enquote {\bibinfo {title} {{Shear thickening in highly viscous granular
  suspensions}},}\ }\href {\doibase 10.1209/0295-5075/107/68004} {\bibfield
  {journal} {\bibinfo  {journal} {Europhysics Letters}\ }\textbf {\bibinfo
  {volume} {107}},\ \bibinfo {pages} {68004} (\bibinfo {year}
  {2014})}\BibitemShut {NoStop}%
\bibitem [{\citenamefont {Lin}\ \emph {et~al.}(2015)\citenamefont {Lin},
  \citenamefont {Guy}, \citenamefont {Hermes}, \citenamefont {Ness},
  \citenamefont {Sun}, \citenamefont {Poon},\ and\ \citenamefont
  {Cohen}}]{Lin2015}%
  \BibitemOpen
  \bibfield  {author} {\bibinfo {author} {\bibfnamefont {N.~Y.~C.}\
  \bibnamefont {Lin}}, \bibinfo {author} {\bibfnamefont {B.~M.}\ \bibnamefont
  {Guy}}, \bibinfo {author} {\bibfnamefont {M.}~\bibnamefont {Hermes}},
  \bibinfo {author} {\bibfnamefont {C.}~\bibnamefont {Ness}}, \bibinfo {author}
  {\bibfnamefont {J.}~\bibnamefont {Sun}}, \bibinfo {author} {\bibfnamefont
  {W.~C.~K.}\ \bibnamefont {Poon}}, \ and\ \bibinfo {author} {\bibfnamefont
  {I.}~\bibnamefont {Cohen}},\ }\bibfield  {title} {\enquote {\bibinfo {title}
  {{Hydrodynamic and Contact Contributions to Continuous Shear Thickening in
  Colloidal Suspensions}},}\ }\href {\doibase 10.1103/PhysRevLett.115.228304}
  {\bibfield  {journal} {\bibinfo  {journal} {Physical Review Letters}\
  }\textbf {\bibinfo {volume} {115}},\ \bibinfo {pages} {1--5} (\bibinfo {year}
  {2015})}\BibitemShut {NoStop}%
\bibitem [{\citenamefont {Fernandez}\ \emph {et~al.}(2013)\citenamefont
  {Fernandez}, \citenamefont {Mani}, \citenamefont {Rinaldi}, \citenamefont
  {Kadau}, \citenamefont {Mosquet}, \citenamefont {Lombois-Burger},
  \citenamefont {Cayer-Barrioz}, \citenamefont {Herrmann}, \citenamefont
  {Spencer},\ and\ \citenamefont {Isa}}]{Fernandez_2013}%
  \BibitemOpen
  \bibfield  {author} {\bibinfo {author} {\bibfnamefont {N.}~\bibnamefont
  {Fernandez}}, \bibinfo {author} {\bibfnamefont {R.}~\bibnamefont {Mani}},
  \bibinfo {author} {\bibfnamefont {D.}~\bibnamefont {Rinaldi}}, \bibinfo
  {author} {\bibfnamefont {D.}~\bibnamefont {Kadau}}, \bibinfo {author}
  {\bibfnamefont {M.}~\bibnamefont {Mosquet}}, \bibinfo {author} {\bibfnamefont
  {H.}~\bibnamefont {Lombois-Burger}}, \bibinfo {author} {\bibfnamefont
  {J.}~\bibnamefont {Cayer-Barrioz}}, \bibinfo {author} {\bibfnamefont {H.~J.}\
  \bibnamefont {Herrmann}}, \bibinfo {author} {\bibfnamefont {N.~D.}\
  \bibnamefont {Spencer}}, \ and\ \bibinfo {author} {\bibfnamefont
  {L.}~\bibnamefont {Isa}},\ }\bibfield  {title} {\enquote {\bibinfo {title}
  {Microscopic mechanism for shear thickening of non-{B}rownian suspensions},}\
  }\href@noop {} {\bibfield  {journal} {\bibinfo  {journal} {Phys. Rev. Lett.}\
  }\textbf {\bibinfo {volume} {111}},\ \bibinfo {pages} {108301} (\bibinfo
  {year} {2013})}\BibitemShut {NoStop}%
\bibitem [{\citenamefont {Mari}\ \emph {et~al.}(2014)\citenamefont {Mari},
  \citenamefont {Seto}, \citenamefont {Morris},\ and\ \citenamefont
  {Denn}}]{Mari_2014}%
  \BibitemOpen
  \bibfield  {author} {\bibinfo {author} {\bibfnamefont {R.}~\bibnamefont
  {Mari}}, \bibinfo {author} {\bibfnamefont {R.}~\bibnamefont {Seto}}, \bibinfo
  {author} {\bibfnamefont {J.~F.}\ \bibnamefont {Morris}}, \ and\ \bibinfo
  {author} {\bibfnamefont {M.~M.}\ \bibnamefont {Denn}},\ }\bibfield  {title}
  {\enquote {\bibinfo {title} {Shear thickening, frictionless and frictional
  rheologies in non-{B}rownian suspensions},}\ }\href@noop {} {\bibfield
  {journal} {\bibinfo  {journal} {J. Rheol.}\ }\textbf {\bibinfo {volume}
  {58}},\ \bibinfo {pages} {1693--1724} (\bibinfo {year} {2014})}\BibitemShut
  {NoStop}%
\bibitem [{\citenamefont {Guy}, \citenamefont {Hermes},\ and\ \citenamefont
  {Poon}(2015)}]{Guy_2015}%
  \BibitemOpen
  \bibfield  {author} {\bibinfo {author} {\bibfnamefont {B.~M.}\ \bibnamefont
  {Guy}}, \bibinfo {author} {\bibfnamefont {M.}~\bibnamefont {Hermes}}, \ and\
  \bibinfo {author} {\bibfnamefont {W.~C.~K.}\ \bibnamefont {Poon}},\
  }\bibfield  {title} {\enquote {\bibinfo {title} {Towards a unified
  description of the rheology of hard-particle suspensions},}\ }\href@noop {}
  {\bibfield  {journal} {\bibinfo  {journal} {Phys. Rev. Lett.}\ }\textbf
  {\bibinfo {volume} {115}},\ \bibinfo {pages} {088304} (\bibinfo {year}
  {2015})}\BibitemShut {NoStop}%
\bibitem [{\citenamefont {Ness}\ and\ \citenamefont {Sun}(2016)}]{Ness_2016}%
  \BibitemOpen
  \bibfield  {author} {\bibinfo {author} {\bibfnamefont {C.}~\bibnamefont
  {Ness}}\ and\ \bibinfo {author} {\bibfnamefont {J.}~\bibnamefont {Sun}},\
  }\bibfield  {title} {\enquote {\bibinfo {title} {Shear thickening regimes of
  dense non-{B}rownian suspensions},}\ }\href@noop {} {\bibfield  {journal}
  {\bibinfo  {journal} {Soft Matter}\ }\textbf {\bibinfo {volume} {12}},\
  \bibinfo {pages} {914--924} (\bibinfo {year} {2016})}\BibitemShut {NoStop}%
\bibitem [{\citenamefont {Clavaud}\ \emph {et~al.}(2017)\citenamefont
  {Clavaud}, \citenamefont {B{\'e}rut}, \citenamefont {Metzger},\ and\
  \citenamefont {Forterre}}]{Clavaud_2017}%
  \BibitemOpen
  \bibfield  {author} {\bibinfo {author} {\bibfnamefont {C.}~\bibnamefont
  {Clavaud}}, \bibinfo {author} {\bibfnamefont {A.}~\bibnamefont {B{\'e}rut}},
  \bibinfo {author} {\bibfnamefont {B.}~\bibnamefont {Metzger}}, \ and\
  \bibinfo {author} {\bibfnamefont {Y.}~\bibnamefont {Forterre}},\ }\bibfield
  {title} {\enquote {\bibinfo {title} {Revealing the frictional transition in
  shear-thickening suspensions},}\ }\href@noop {} {\bibfield  {journal}
  {\bibinfo  {journal} {Proc. Natl. Acad. Sci. U.S.A.}\ ,\ \bibinfo {pages}
  {5147--5152}} (\bibinfo {year} {2017})}\BibitemShut {NoStop}%
\bibitem [{\citenamefont {Comtet}\ \emph {et~al.}(2017)\citenamefont {Comtet},
  \citenamefont {Chatt{\'e}}, \citenamefont {Nigu{\`e}s}, \citenamefont
  {Bocquet}, \citenamefont {Siria},\ and\ \citenamefont {Colin}}]{Comtet_2017}%
  \BibitemOpen
  \bibfield  {author} {\bibinfo {author} {\bibfnamefont {J.}~\bibnamefont
  {Comtet}}, \bibinfo {author} {\bibfnamefont {G.}~\bibnamefont {Chatt{\'e}}},
  \bibinfo {author} {\bibfnamefont {A.}~\bibnamefont {Nigu{\`e}s}}, \bibinfo
  {author} {\bibfnamefont {L.}~\bibnamefont {Bocquet}}, \bibinfo {author}
  {\bibfnamefont {A.}~\bibnamefont {Siria}}, \ and\ \bibinfo {author}
  {\bibfnamefont {A.}~\bibnamefont {Colin}},\ }\bibfield  {title} {\enquote
  {\bibinfo {title} {Pairwise frictional profile between particles determines
  discontinuous shear thickening transition in non-colloidal suspensions.}}\
  }\href@noop {} {\bibfield  {journal} {\bibinfo  {journal} {Nat. Comm.}\
  }\textbf {\bibinfo {volume} {8}},\ \bibinfo {pages} {15633} (\bibinfo {year}
  {2017})}\BibitemShut {NoStop}%
\bibitem [{\citenamefont {Lootens}\ \emph {et~al.}(2005)\citenamefont
  {Lootens}, \citenamefont {van Damme}, \citenamefont {H{\'{e}}mar},\ and\
  \citenamefont {H{\'{e}}braud}}]{Lootens2005}%
  \BibitemOpen
  \bibfield  {author} {\bibinfo {author} {\bibfnamefont {D.}~\bibnamefont
  {Lootens}}, \bibinfo {author} {\bibfnamefont {H.}~\bibnamefont {van Damme}},
  \bibinfo {author} {\bibfnamefont {Y.}~\bibnamefont {H{\'{e}}mar}}, \ and\
  \bibinfo {author} {\bibfnamefont {P.}~\bibnamefont {H{\'{e}}braud}},\
  }\bibfield  {title} {\enquote {\bibinfo {title} {{Dilatant Flow of
  Concentrated Suspensions of Rough Particles}},}\ }\href {\doibase
  10.1103/PhysRevLett.95.268302} {\bibfield  {journal} {\bibinfo  {journal}
  {Physical Review Letters}\ }\textbf {\bibinfo {volume} {95}},\ \bibinfo
  {pages} {268302} (\bibinfo {year} {2005})}\BibitemShut {NoStop}%
\bibitem [{\citenamefont {Hsu}\ \emph {et~al.}(2018)\citenamefont {Hsu},
  \citenamefont {Ramakrishna}, \citenamefont {Zanini}, \citenamefont
  {Spencer},\ and\ \citenamefont {Isa}}]{Hsu_2018}%
  \BibitemOpen
  \bibfield  {author} {\bibinfo {author} {\bibfnamefont {C.-P.}\ \bibnamefont
  {Hsu}}, \bibinfo {author} {\bibfnamefont {S.~N.}\ \bibnamefont
  {Ramakrishna}}, \bibinfo {author} {\bibfnamefont {M.}~\bibnamefont {Zanini}},
  \bibinfo {author} {\bibfnamefont {N.~D.}\ \bibnamefont {Spencer}}, \ and\
  \bibinfo {author} {\bibfnamefont {L.}~\bibnamefont {Isa}},\ }\bibfield
  {title} {\enquote {\bibinfo {title} {Roughness-dependent tribology effects on
  discontinuous shear thickening},}\ }\href@noop {} {\bibfield  {journal}
  {\bibinfo  {journal} {Proc. Nat. Acad. Sci.}\ } (\bibinfo {year}
  {2018})}\BibitemShut {NoStop}%
\bibitem [{\citenamefont {James}\ \emph {et~al.}(2018)\citenamefont {James},
  \citenamefont {Han}, \citenamefont {de~la Cruz}, \citenamefont {Jureller},\
  and\ \citenamefont {Jaeger}}]{james2018interparticle}%
  \BibitemOpen
  \bibfield  {author} {\bibinfo {author} {\bibfnamefont {N.~M.}\ \bibnamefont
  {James}}, \bibinfo {author} {\bibfnamefont {E.}~\bibnamefont {Han}}, \bibinfo
  {author} {\bibfnamefont {R.~A.~L.}\ \bibnamefont {de~la Cruz}}, \bibinfo
  {author} {\bibfnamefont {J.}~\bibnamefont {Jureller}}, \ and\ \bibinfo
  {author} {\bibfnamefont {H.~M.}\ \bibnamefont {Jaeger}},\ }\bibfield  {title}
  {\enquote {\bibinfo {title} {Interparticle hydrogen bonding can elicit shear
  jamming in dense suspensions},}\ }\href@noop {} {\bibfield  {journal}
  {\bibinfo  {journal} {Nature materials}\ }\textbf {\bibinfo {volume} {17}},\
  \bibinfo {pages} {965} (\bibinfo {year} {2018})}\BibitemShut {NoStop}%
\bibitem [{\citenamefont {Brown}\ \emph {et~al.}(2010)\citenamefont {Brown},
  \citenamefont {Forman}, \citenamefont {Orellana}, \citenamefont {Hanjun},
  \citenamefont {Maynor}, \citenamefont {Betts}, \citenamefont {DeSimone},\
  and\ \citenamefont {Jaeger}}]{Brown_2010}%
  \BibitemOpen
  \bibfield  {author} {\bibinfo {author} {\bibfnamefont {E.}~\bibnamefont
  {Brown}}, \bibinfo {author} {\bibfnamefont {N.~A.}\ \bibnamefont {Forman}},
  \bibinfo {author} {\bibfnamefont {C.~S.}\ \bibnamefont {Orellana}}, \bibinfo
  {author} {\bibfnamefont {Z.}~\bibnamefont {Hanjun}}, \bibinfo {author}
  {\bibfnamefont {B.~W.}\ \bibnamefont {Maynor}}, \bibinfo {author}
  {\bibfnamefont {D.~E.}\ \bibnamefont {Betts}}, \bibinfo {author}
  {\bibfnamefont {J.~M.}\ \bibnamefont {DeSimone}}, \ and\ \bibinfo {author}
  {\bibfnamefont {H.~M.}\ \bibnamefont {Jaeger}},\ }\bibfield  {title}
  {\enquote {\bibinfo {title} {Generality of shear thickening in dense
  suspensions},}\ }\href@noop {} {\bibfield  {journal} {\bibinfo  {journal}
  {Nat. Mater.}\ }\textbf {\bibinfo {volume} {9}},\ \bibinfo {pages} {220--224}
  (\bibinfo {year} {2010})}\BibitemShut {NoStop}%
\bibitem [{\citenamefont {Singh}\ \emph {et~al.}(2019)\citenamefont {Singh},
  \citenamefont {Pednekar}, \citenamefont {Chun}, \citenamefont {Denn},\ and\
  \citenamefont {Morris}}]{Singh_2019}%
  \BibitemOpen
  \bibfield  {author} {\bibinfo {author} {\bibfnamefont {A.}~\bibnamefont
  {Singh}}, \bibinfo {author} {\bibfnamefont {S.}~\bibnamefont {Pednekar}},
  \bibinfo {author} {\bibfnamefont {J.}~\bibnamefont {Chun}}, \bibinfo {author}
  {\bibfnamefont {M.~M.}\ \bibnamefont {Denn}}, \ and\ \bibinfo {author}
  {\bibfnamefont {J.~F.}\ \bibnamefont {Morris}},\ }\bibfield  {title}
  {\enquote {\bibinfo {title} {From yielding to shear jamming in a cohesive
  frictional suspension},}\ }\href@noop {} {\bibfield  {journal} {\bibinfo
  {journal} {Physical review letters}\ }\textbf {\bibinfo {volume} {122}},\
  \bibinfo {pages} {098004} (\bibinfo {year} {2019})}\BibitemShut {NoStop}%
\bibitem [{\citenamefont {Lootens}, \citenamefont {{Van Damme}},\ and\
  \citenamefont {H{\'{e}}braud}(2003)}]{Lootens2003}%
  \BibitemOpen
  \bibfield  {author} {\bibinfo {author} {\bibfnamefont {D.}~\bibnamefont
  {Lootens}}, \bibinfo {author} {\bibfnamefont {H.}~\bibnamefont {{Van
  Damme}}}, \ and\ \bibinfo {author} {\bibfnamefont {P.}~\bibnamefont
  {H{\'{e}}braud}},\ }\bibfield  {title} {\enquote {\bibinfo {title} {{Giant
  Stress Fluctuations at the Jamming Transition}},}\ }\href {\doibase
  10.1103/PhysRevLett.90.178301} {\bibfield  {journal} {\bibinfo  {journal}
  {Physical Review Letters}\ }\textbf {\bibinfo {volume} {90}},\ \bibinfo
  {pages} {178301} (\bibinfo {year} {2003})}\BibitemShut {NoStop}%
\bibitem [{\citenamefont {Dasan}\ \emph {et~al.}(2002)\citenamefont {Dasan},
  \citenamefont {Ramamohan}, \citenamefont {Singh},\ and\ \citenamefont
  {Nott}}]{Dasan2002}%
  \BibitemOpen
  \bibfield  {author} {\bibinfo {author} {\bibfnamefont {J.}~\bibnamefont
  {Dasan}}, \bibinfo {author} {\bibfnamefont {T.}~\bibnamefont {Ramamohan}},
  \bibinfo {author} {\bibfnamefont {A.}~\bibnamefont {Singh}}, \ and\ \bibinfo
  {author} {\bibfnamefont {P.}~\bibnamefont {Nott}},\ }\bibfield  {title}
  {\enquote {\bibinfo {title} {{Stress fluctuations in sheared Stokesian
  suspensions}},}\ }\href {\doibase 10.1103/PhysRevE.66.021409} {\bibfield
  {journal} {\bibinfo  {journal} {Physical Review E}\ }\textbf {\bibinfo
  {volume} {66}},\ \bibinfo {pages} {021409} (\bibinfo {year}
  {2002})}\BibitemShut {NoStop}%
\bibitem [{\citenamefont {Miller}, \citenamefont {O'Hern},\ and\ \citenamefont
  {Behringer}(1996)}]{MillerBrian;OHernCorey;Behringer1996}%
  \BibitemOpen
  \bibfield  {author} {\bibinfo {author} {\bibfnamefont {B.}~\bibnamefont
  {Miller}}, \bibinfo {author} {\bibfnamefont {C.}~\bibnamefont {O'Hern}}, \
  and\ \bibinfo {author} {\bibfnamefont {R.~P.}\ \bibnamefont {Behringer}},\
  }\bibfield  {title} {\enquote {\bibinfo {title} {{Stress fluctuations in
  continuously sheared dense granular materials.}}}\ }\href
  {http://journals.aps.org/prl/abstract/10.1103/PhysRevLett.77.3110} {\bibfield
   {journal} {\bibinfo  {journal} {Physical Revew Letters}\ }\textbf {\bibinfo
  {volume} {77}},\ \bibinfo {pages} {3110--3113} (\bibinfo {year}
  {1996})}\BibitemShut {NoStop}%
\bibitem [{\citenamefont {Howell}, \citenamefont {Behringer},\ and\
  \citenamefont {Veje}()}]{Howell1999a}%
  \BibitemOpen
  \bibfield  {author} {\bibinfo {author} {\bibfnamefont {D.}~\bibnamefont
  {Howell}}, \bibinfo {author} {\bibfnamefont {R.~P.}\ \bibnamefont
  {Behringer}}, \ and\ \bibinfo {author} {\bibfnamefont {C.}~\bibnamefont
  {Veje}},\ }\bibfield  {title} {\enquote {\bibinfo {title} {{Stress
  fluctuations in a 2D granular Couette experiment: a continuous
  transition}},}\ }\href
  {http://journals.aps.org/prl/abstract/10.1103/PhysRevLett.82.5241} {\bibfield
   {journal} {\bibinfo  {journal} {Physical Review Letters}\ }\textbf {\bibinfo
  {volume} {82}},\ \bibinfo {pages} {5241--5244}}\BibitemShut {NoStop}%
\bibitem [{\citenamefont {Blair}\ \emph {et~al.}(2001)\citenamefont {Blair},
  \citenamefont {Mueggenburg}, \citenamefont {Marshall}, \citenamefont
  {Jaeger},\ and\ \citenamefont {Nagel}}]{Blair2001}%
  \BibitemOpen
  \bibfield  {author} {\bibinfo {author} {\bibfnamefont {D.}~\bibnamefont
  {Blair}}, \bibinfo {author} {\bibfnamefont {N.}~\bibnamefont {Mueggenburg}},
  \bibinfo {author} {\bibfnamefont {A.}~\bibnamefont {Marshall}}, \bibinfo
  {author} {\bibfnamefont {H.}~\bibnamefont {Jaeger}}, \ and\ \bibinfo {author}
  {\bibfnamefont {S.}~\bibnamefont {Nagel}},\ }\bibfield  {title} {\enquote
  {\bibinfo {title} {{Force distributions in three-dimensional granular
  assemblies: Effects of packing order and interparticle friction}},}\ }\href
  {\doibase 10.1103/PhysRevE.63.041304} {\bibfield  {journal} {\bibinfo
  {journal} {Physical Review E}\ }\textbf {\bibinfo {volume} {63}},\ \bibinfo
  {pages} {041304} (\bibinfo {year} {2001})}\BibitemShut {NoStop}%
\bibitem [{\citenamefont {Corwin}, \citenamefont {Jaeger},\ and\ \citenamefont
  {Nagel}(2005)}]{Corwin2005}%
  \BibitemOpen
  \bibfield  {author} {\bibinfo {author} {\bibfnamefont {E.~I.}\ \bibnamefont
  {Corwin}}, \bibinfo {author} {\bibfnamefont {H.~M.}\ \bibnamefont {Jaeger}},
  \ and\ \bibinfo {author} {\bibfnamefont {S.~R.}\ \bibnamefont {Nagel}},\
  }\bibfield  {title} {\enquote {\bibinfo {title} {{Structural signature of
  jamming in granular media.}}}\ }\href {\doibase 10.1038/nature03698}
  {\bibfield  {journal} {\bibinfo  {journal} {Nature}\ }\textbf {\bibinfo
  {volume} {435}},\ \bibinfo {pages} {1075--8} (\bibinfo {year}
  {2005})}\BibitemShut {NoStop}%
\bibitem [{\citenamefont {Corwin}\ \emph {et~al.}(2008)\citenamefont {Corwin},
  \citenamefont {Hoke}, \citenamefont {Jaeger},\ and\ \citenamefont
  {Nagel}}]{Corwin2008}%
  \BibitemOpen
  \bibfield  {author} {\bibinfo {author} {\bibfnamefont {E.~I.}\ \bibnamefont
  {Corwin}}, \bibinfo {author} {\bibfnamefont {E.~T.}\ \bibnamefont {Hoke}},
  \bibinfo {author} {\bibfnamefont {H.~M.}\ \bibnamefont {Jaeger}}, \ and\
  \bibinfo {author} {\bibfnamefont {S.~R.}\ \bibnamefont {Nagel}},\ }\bibfield
  {title} {\enquote {\bibinfo {title} {{Temporal force fluctuations measured by
  tracking individual particles in granular materials under shear}},}\ }\href
  {\doibase 10.1103/PhysRevE.77.061308} {\bibfield  {journal} {\bibinfo
  {journal} {Physical Review E}\ }\textbf {\bibinfo {volume} {77}},\ \bibinfo
  {pages} {061308} (\bibinfo {year} {2008})}\BibitemShut {NoStop}%
\bibitem [{\citenamefont {Singh}\ \emph {et~al.}(2014)\citenamefont {Singh},
  \citenamefont {Magnanimo}, \citenamefont {Saitoh},\ and\ \citenamefont
  {Luding}}]{singh2014effect}%
  \BibitemOpen
  \bibfield  {author} {\bibinfo {author} {\bibfnamefont {A.}~\bibnamefont
  {Singh}}, \bibinfo {author} {\bibfnamefont {V.}~\bibnamefont {Magnanimo}},
  \bibinfo {author} {\bibfnamefont {K.}~\bibnamefont {Saitoh}}, \ and\ \bibinfo
  {author} {\bibfnamefont {S.}~\bibnamefont {Luding}},\ }\bibfield  {title}
  {\enquote {\bibinfo {title} {Effect of cohesion on shear banding in
  quasistatic granular materials},}\ }\href@noop {} {\bibfield  {journal}
  {\bibinfo  {journal} {Physical Review E}\ }\textbf {\bibinfo {volume} {90}},\
  \bibinfo {pages} {022202} (\bibinfo {year} {2014})}\BibitemShut {NoStop}%
\bibitem [{\citenamefont {Singh}\ \emph {et~al.}(2018)\citenamefont {Singh},
  \citenamefont {Mari}, \citenamefont {Denn},\ and\ \citenamefont
  {Morris}}]{Singh_2018}%
  \BibitemOpen
  \bibfield  {author} {\bibinfo {author} {\bibfnamefont {A.}~\bibnamefont
  {Singh}}, \bibinfo {author} {\bibfnamefont {R.}~\bibnamefont {Mari}},
  \bibinfo {author} {\bibfnamefont {M.~M.}\ \bibnamefont {Denn}}, \ and\
  \bibinfo {author} {\bibfnamefont {J.~F.}\ \bibnamefont {Morris}},\ }\bibfield
   {title} {\enquote {\bibinfo {title} {A constitutive model for simple shear
  of dense frictional suspensions},}\ }\href@noop {} {\bibfield  {journal}
  {\bibinfo  {journal} {J. Rheol.}\ }\textbf {\bibinfo {volume} {62}},\
  \bibinfo {pages} {457--468} (\bibinfo {year} {2018})}\BibitemShut {NoStop}%
\bibitem [{\citenamefont {Royer}, \citenamefont {Blair},\ and\ \citenamefont
  {Hudson}(2016)}]{Royer_2016}%
  \BibitemOpen
  \bibfield  {author} {\bibinfo {author} {\bibfnamefont {J.~R.}\ \bibnamefont
  {Royer}}, \bibinfo {author} {\bibfnamefont {D.~L.}\ \bibnamefont {Blair}}, \
  and\ \bibinfo {author} {\bibfnamefont {S.~D.}\ \bibnamefont {Hudson}},\
  }\bibfield  {title} {\enquote {\bibinfo {title} {Rheological signature of
  frictional interactions in shear thickening suspensions},}\ }\href@noop {}
  {\bibfield  {journal} {\bibinfo  {journal} {Phys. Rev. Lett.}\ }\textbf
  {\bibinfo {volume} {116}},\ \bibinfo {pages} {188301} (\bibinfo {year}
  {2016})}\BibitemShut {NoStop}%
\bibitem [{\citenamefont {Boersma}, \citenamefont {Laven},\ and\ \citenamefont
  {Stein}(1990)}]{boersma_1990}%
  \BibitemOpen
  \bibfield  {author} {\bibinfo {author} {\bibfnamefont {W.~H.}\ \bibnamefont
  {Boersma}}, \bibinfo {author} {\bibfnamefont {J.}~\bibnamefont {Laven}}, \
  and\ \bibinfo {author} {\bibfnamefont {H.~N.}\ \bibnamefont {Stein}},\
  }\bibfield  {title} {\enquote {\bibinfo {title} {Shear thickening (dilatancy)
  in concentrated dispersions},}\ }\href@noop {} {\bibfield  {journal}
  {\bibinfo  {journal} {AIChE J.}\ }\textbf {\bibinfo {volume} {36}},\ \bibinfo
  {pages} {321--332} (\bibinfo {year} {1990})}\BibitemShut {NoStop}%
\bibitem [{\citenamefont {Rathee}, \citenamefont {Blair},\ and\ \citenamefont
  {Urbach}(2017)}]{rathee_2017}%
  \BibitemOpen
  \bibfield  {author} {\bibinfo {author} {\bibfnamefont {V.}~\bibnamefont
  {Rathee}}, \bibinfo {author} {\bibfnamefont {D.~L.}\ \bibnamefont {Blair}}, \
  and\ \bibinfo {author} {\bibfnamefont {J.~S.}\ \bibnamefont {Urbach}},\
  }\bibfield  {title} {\enquote {\bibinfo {title} {Localized stress
  fluctuations drive shear thickening in dense suspensions},}\ }\href@noop {}
  {\bibfield  {journal} {\bibinfo  {journal} {Proceedings of the National
  Academy of Sciences}\ }\textbf {\bibinfo {volume} {114}},\ \bibinfo {pages}
  {8740--8745} (\bibinfo {year} {2017})}\BibitemShut {NoStop}%
\bibitem [{\citenamefont {Coppersmith}\ \emph {et~al.}(1996)\citenamefont
  {Coppersmith}, \citenamefont {Liu}, \citenamefont {Majumdar}, \citenamefont
  {Narayan},\ and\ \citenamefont {Witten}}]{Coppersmith1996}%
  \BibitemOpen
  \bibfield  {author} {\bibinfo {author} {\bibfnamefont {S.}~\bibnamefont
  {Coppersmith}}, \bibinfo {author} {\bibfnamefont {C.}~\bibnamefont {Liu}},
  \bibinfo {author} {\bibfnamefont {S.}~\bibnamefont {Majumdar}}, \bibinfo
  {author} {\bibfnamefont {O.}~\bibnamefont {Narayan}}, \ and\ \bibinfo
  {author} {\bibfnamefont {T.}~\bibnamefont {Witten}},\ }\bibfield  {title}
  {\enquote {\bibinfo {title} {{Model for force fluctuations in bead packs.}}}\
  }\href {http://www.ncbi.nlm.nih.gov/pubmed/9964795} {\bibfield  {journal}
  {\bibinfo  {journal} {Physical review. E}\ }\textbf {\bibinfo {volume}
  {53}},\ \bibinfo {pages} {4673--4685} (\bibinfo {year} {1996})}\BibitemShut
  {NoStop}%
\bibitem [{\citenamefont {Bi}\ \emph {et~al.}(2011)\citenamefont {Bi},
  \citenamefont {Zhang}, \citenamefont {Chakraborty},\ and\ \citenamefont
  {Behringer}}]{Bi2011}%
  \BibitemOpen
  \bibfield  {author} {\bibinfo {author} {\bibfnamefont {D.}~\bibnamefont
  {Bi}}, \bibinfo {author} {\bibfnamefont {J.}~\bibnamefont {Zhang}}, \bibinfo
  {author} {\bibfnamefont {B.}~\bibnamefont {Chakraborty}}, \ and\ \bibinfo
  {author} {\bibfnamefont {R.~P.}\ \bibnamefont {Behringer}},\ }\bibfield
  {title} {\enquote {\bibinfo {title} {Jamming by shear},}\ }\href {\doibase
  10.1038/nature10667} {\bibfield  {journal} {\bibinfo  {journal} {Nature}\
  }\textbf {\bibinfo {volume} {480}},\ \bibinfo {pages} {355--358} (\bibinfo
  {year} {2011})}\BibitemShut {NoStop}%
\end{thebibliography}%

\end{document}